\documentclass{jpsj2}
\usepackage{graphicx}      % Include figure files
\usepackage{dcolumn}       % Align table columns on decimal point
\usepackage{bm}            % bold math
\usepackage{rsfs}          % calligraphic math character {\mathscr H}

\newcommand{\bra}{\langle}
\newcommand{\ket}{\rangle}
\title{Generation and Suppression of Decoherence  
\\ in Artificial Environment for Qubit System}

\author{
 Yasushi \textsc{Kondo}$^{1}$
    \thanks{E-mail address: kondo@phys.kindai.ac.jp}, 
 Mikio \textsc{Nakahara}$^{1}$
    \thanks{Also at Low Temperature Laboratory, 
             Helsinki University of Technology, Box 2200, 
             02015 TKK, Finland} 
 Shogo   \textsc{Tanimura}$^{2}$,
 Sachiko \textsc{Kitajima}$^{3}$, \\
 Chikako \textsc{Uchiyama}$^{4}$,
and 
 Fumiaki \textsc{Shibata}$^{5}$
}

\inst{
$^{1}$Department of Physics, Kinki University, Higashi-Osaka 577-8502, Japan \\
$^{2}$Graduate School of Informatics, Kyoto University, Kyoto 606-8501, Japan \\
$^{3}$Graduate School of Pure and Applied Sciences, 
      University of Tsukuba, \\
      Tenodai 1-1-1 Tsukuba, Ibaraki 305-8571, Japan \\
$^{4}$Faculty of Engineering, University of Yamanashi, 
4-3-11 Takeda, Kofu, Yamanashi 400-8511, Japan 
$^{5}$Department of Physics, Faculty of Science, 
Ochanomizu University, Tokyo 112-8610, Japan 
}

\abst{
A quantum system with finite degrees of freedom can 
simulate a composite of a system and its environment 
if the state of the hypothetical environment 
is randomized by external manipulation.
We propose to examine various techniques 
in quantum information processing in virtual noisy environment
with two concrete examples.  
One simulates phase decoherence of a single qubit 
in a transmission line, and the other does in a quantum memory. 
In both cases, the bang-bang control, a typical example of 
useful techniques employed in quantum information processing, is observed 
to be effective to suppress decoherence. 
}

\kword{NMR quantum computer, channel, decoherence, bang-bang control}
\begin{document}
\maketitle

%=====================================================================
\section{Introduction} 
%=====================================================================
Quantum computing currently attracts a lot of attention 
since it is expected to solve some of computationally hard problems 
for a conventional digital computer~\cite{ref:1}. 
Numerous realizations of a quantum computer have been proposed 
to date. Among others, a liquid-state nuclear magnetic resonance
(NMR)
quantum computer is regarded as most successful. 
Demonstration of Shor's factorization algorithm
with NMR is one of the most remarkable achievements~\cite{VSB01}.
Liquid state NMR will be denoted simply as NMR throughout this paper.

Although the current NMR quantum computer 
is suspected not to be a true quantum computer 
because of its poor spin polarization 
at room temperature~\cite{pt}, 
it still works as a test bench for a more realistic
quantum computer. 
For example, 
we have demonstrated our time-optimal implementation 
of two-qubit quantum algorithms by
using an NMR quantum computer~\cite{qaa,warp}.

A molecule employed in 
NMR experiments can be arranged to work not only as
a quantum register but also as a composite system of 
a quantum system and its environment. 
It is possible to introduce decoherence phenomena in the quantum subsystem 
by manipulating the environment subsystem. 
Moreover the combined system
can be employed as a test bench to develop techniques to protect a
quantum system from decoherence.
Decoherence is one of the primary obstacles in constructing
a working quantum computer
and must be suppressed somehow.
Decoherence effect in a quantum register has also been studied 
elsewhere~\cite{Zanardi}.

The main purpose of this paper is two-fold. 
Firstly, we show that 
decoherence can be generated by manipulating the artificial environment.
Secondly, we verify by NMR experiments
that decoherence control methods,
such as a bang-bang control~\cite{bb,uchiyama}, actually
suppress decoherence.
It should be noted that demonstration of the effectiveness of
decoherence control methods is rather difficult in other systems 
due to their extremely short coherence times.

Section \ref{sec:decoherence} is a brief review of the theory of a
quantum channel,
which is a useful formalism to describe decoherence in a general context.
In Section \ref{sec:one-qubit}
we describe decoherence of a one-qubit system in terms of a
quantum channel.
There we discuss the method to suppress decoherence by a bang-bang control.
In Section \ref{sec:two-qubit}
we show that a two-qubit system may be regarded as a composite
of a system (qubit~1) and an environment (qubit~2).
We introduce an artificial environment by manipulating qubit~2,
which causes decoherence in qubit~1. 
We show two illustrating examples 
and calculate decoherence rates in these environments.
We also show that 
application of a bang-bang control to qubit~1 suppresses decoherence
in both cases.
In Section \ref{sec:experiment} we report the results of our experiments, 
which support our theory.
Using a two-spin molecule we demonstrate
the generation of decoherence
and its suppression by the bang-bang control.
Section \ref{sec:conclusion} is devoted to summary and conclusions.

%=====================================================================
\section{Decoherence}
%=====================================================================
\label{sec:decoherence}
\subsection{Quantum Channel}

Decoherence is an irreversible change of a state of a quantum system
which has quantum correlation with its environment.
The change of the state of the system becomes irreversible 
due to our lack of knowledge about the state of the environment.

Decoherence is formulated in terms of a
channel or a quantum operation~\cite{Hellwig,Kraus,ref:1,ECGS61}
as follows.
Let $ {\mathscr H}_{\rm s} $ and $ {\mathscr H}_{\rm e} $
be the Hilbert spaces of the system and 
the environment, respectively.
The initial state of the system is represented by 
the density matrix $ \rho_{\rm s} $
while that of the environment by $ \rho_{\rm e} $.
The state of the whole system changes
following the time-evolution law,
\begin{equation}
	\rho_{\rm s} \otimes \rho_{\rm e}
	\; \to \;
	U \rho_{\rm s} \otimes \rho_{\rm e} \, U^\dagger.
	\label{unitary time-evolution law}
\end{equation}
Here $ U $ is a unitary operator acting on the Hilbert space
of the composite system
$ {\mathscr H}_{\rm s} \otimes {\mathscr H}_{\rm e} $.
%%%%%%%% resposnes to the referee %%%%%%%%%%
% We consider the case in which it is impossible to write $U$
% as $U_{\rm s} \otimes U_{\rm e}$,
% where $U_{\rm s}$ and $U_{\rm e}$ act only on the system 
% and environment, respectively.
We consider the case in which the initial state is a separable state
$ \rho = \rho_{\rm s} \otimes \rho_{\rm e} $.
%%%%%%%%%%%%%%%%%%%%%%%%%%%%%%%%%%%%%%%%%%%%
The states of the system and the environment are correlated via
the transformation~(\ref{unitary time-evolution law}).
Needless to say, 
the unitary transformation~(\ref{unitary time-evolution law})
is a reversible change.
If we are interested only in the state of the system,
the measurement outcomes are completely described
by the reduced density matrix
\begin{equation}
	\rho'_{\rm s} 
	= {\mathscr E} ( \rho_{\rm s} )
	=
	{\rm Tr}_{\rm e} 
	\Big(
		U
		\rho_{\rm s} \otimes \rho_{\rm e} \,
		U^\dagger
	\Big),
	\label{channel}
\end{equation}
where the symbol $ {\rm Tr}_{\rm e} $ denotes
the partial trace over $  {\mathscr H}_{\rm e} $.
The partial trace operation is non-invertible
and the associated loss of information leads to decoherence.
Even if the initial state $ \rho_{\rm s} $ is a pure state,
the transformed state $ {\mathscr E} (\rho_{\rm s}) $
becomes a mixed state in general.

The mapping 
$ \rho_{\rm s} \to {\mathscr E} ( \rho_{\rm s} ) $
is called a channel or a quantum operation~\cite{ref:1}.
A channel is the most general mathematical device to describe
changes of a quantum state, 
including unitary time-evolution, measurement process, decoherence 
and so on.
It is known that 
for a channel 
there is a set of operators $ \{ E_k \} $ acting on $ {\mathscr H}_{\rm s} $
such that
\begin{eqnarray}
&&	{\mathscr E} ( \rho_{\rm s} )
	=
	\sum_{k} E_k \rho_{\rm s} E_k^\dagger,
	\label{operator-sum} \\
&&	\sum_{k} E_k^\dagger E_k = I_{\rm s}.
	\label{trace preserving}
\end{eqnarray}
Here $ I_{\rm s} $ is the identity operator on $ {\mathscr H}_{\rm s} $.
Equation~(\ref{operator-sum}) is called an operator-sum representation
of the channel $ {\mathscr E} $.
Equation~(\ref{trace preserving}) implies 
$ {\rm Tr}_{\rm s} \, {\mathscr E} ( \rho_{\rm s} )
= {\rm Tr}_{\rm s} \, \rho_{\rm s} $, ${\rm Tr}_{\rm s}$ being the trace
over ${\mathscr H}_{\rm s}$, 
and hence it is called the trace-preserving condition.

\subsection{Mixing process as a quantum channel}

There is another approach to defining channels
without resort to partial trace over the Hilbert space of environment.
Assume that we have a set of unitary operators $ \{ U_k \} $,
which act on $ {\mathscr H}_{\rm s} $,
and that we have a set of real numbers
$ \{ p_k \} $ such that $ 0 \le p_k \le 1 $ and $ \sum_k p_k = 1 $.
We then define a transformation of the system density matrix $ \rho_{\rm s} $
by
\begin{equation}
	\rho_{\rm s} 
	\; \to \;
	{\mathscr M} ( \rho_{\rm s} )
	=
	\sum_k p_k \,
	U_k \, \rho_{\rm s} \, U_k^\dagger.
	\label{mixing}
\end{equation}
They satisfy the condition (\ref{trace preserving})
if we put $ E_k = \sqrt{p_k} \, U_k $.
This argument tells us that 
if we apply a set of time-evolution unitary operators
$ \{ U_k \} $ on the system
with a probability distribution $ \{ p_k \} $,
we will observe a decoherence-like phenomenon 
after taking an average of the measured data over $k$.
We call the transformation $ {\mathscr M} $ a mixing process.

Although a mixing process is defined superficially
without referring to an environment,
it is mathematically a special case of 
a channel that is defined through interaction 
between a system and an environment.
For given sets of
unitary operators $ \{ U_k \} $
and probabilities $ \{ p_k \} $,
we can construct a Hilbert space
$ {\mathscr H}_{\rm e} = \{ \, \sum_k c_k | k \ket \, \} $
by demanding formally that $ \{ | k \ket \} $ is a complete orthonormal set.
Moreover, we define an environment density matrix
\begin{equation}
	\rho_{\rm e} = \sum_k p_k | k \ket \bra k |
\end{equation}
and define a unitary operator $ U = \sum_k U_k \otimes |k \rangle \langle
k|$
that acts on
$ {\mathscr H}_{\rm s} \otimes {\mathscr H}_{\rm e} $
as
\begin{equation}
	U \Big( | \psi_{\rm s} \ket \otimes | k \ket \Big)
	=
	\Big( U_k | \psi_{\rm s} \ket \Big) \otimes | k \ket.
\end{equation}
By substituting them into the defining equation
of a channel (\ref{channel}),
we obtain the mixing process (\ref{mixing}).
In this paper we use the mixing process 
as a procedure to build a channel.

%=====================================================================
\section{Decoherence in one-qubit system}
%=====================================================================
\label{sec:one-qubit}

\subsection{Phase flip channel}
Here we introduce the phase flip channel,
which is a typical example of a channel.
We take a one-qubit system and a one-qubit environment for simplicity.
Assume that the initial state of the environment is
\begin{equation}
	| \psi_{\rm e} \ket
	= \sqrt{p}   \, | 0 \ket 
	+ \sqrt{1-p} \, | 1 \ket
	\label{psi_e}
\end{equation}
with a real number $ p $ $ (0 \le p \le 1) $. Here the vectors
$ \{ | 0 \ket,  \, | 1 \ket  \} $ are eigenstates of $ \sigma_z $
such that $ \sigma_z | 0 \ket = | 0 \ket $ and
$ \sigma_z | 1 \ket =-| 1 \ket $, where
$\sigma_{x, y, z}$ are conventional Pauli matrices.
We take a unitary operator
\begin{equation}
	U =I \otimes |0 \rangle \langle 0| + \sigma_z \otimes |1 \rangle
	\langle 1|
	\label{U for flip}
	= 
	\begin{pmatrix}
	1 & 0 & 0 & 0 \\
	0 & 1 & 0 & 0 \\
	0 & 0 & 1 & 0 \\
	0 & 0 & 0 &-1 
	\end{pmatrix}
\end{equation}
which acts on $ {\mathbb C}^2 \otimes {\mathbb C}^2 $, where
$ I $ is the two-dimensional identity matrix.
By substituting them into Eq.~(\ref{channel}) we obtain a channel
\begin{equation}
	{\mathscr E} ( \rho_{\rm s} )
	=
	{\rm Tr}_{\rm e} 
	\Big(
		U
		\rho_{\rm s} \otimes 
		| \psi_{\rm e} \ket \bra \psi_{\rm e} | \,
		U^\dagger
	\Big)
	= E_0 \rho_{\rm s} E_0^\dagger
	+ E_1 \rho_{\rm s} E_1^\dagger
	\label{flip}
\end{equation}
with
\begin{eqnarray}
&&	E_0 
	= \bra 0 | U | \psi_{\rm e} \ket
	= \sqrt{p}\ I, 
\\
&&	E_1 
	= \bra 1 | U | \psi_{\rm e} \ket
	= \sqrt{1-p}\ \sigma_z. 
\end{eqnarray}
Any initial state of a one-qubit system is parametrized as
\begin{equation}
	\rho_{\rm s} 
	=
	\begin{pmatrix}
		\rho_{00} & \rho_{01} \\
		\rho_{10} & \rho_{11} 
	\end{pmatrix}
	= 
	\frac{1}{2}
	( I 
	+ a_x \sigma_x 
	+ a_y \sigma_y 
	+ a_z \sigma_z )
	\label{parameters of rho}
\end{equation}
with real numbers $ a_x, a_y, a_z $ such that
$ a_x^2 + a_y^2 + a_z^2 \le 1 $.
The vector $ (a_x, a_y, a_z) $ is called the Bloch vector.
The angle $ \phi $ defined by
\begin{equation}
	2 \rho_{10} = a_x + i a_y = e^{i \phi} \sqrt{ a_x^2 + a_y^2 } 
	\label{amplitude}
\end{equation}
denotes the azimuthal angle of the Bloch vector,
and is called the phase of the spin.
The complex quantity $ a_x + i a_y $ is called an amplitude
in the context of NMR.
The channel (\ref{flip}) transforms $ \rho_{\rm s} $ to
\begin{eqnarray}
	{\mathscr E} (\rho_{\rm s}) 
&=&	\frac{1}{2} p \,
	( I 
	+ a_x \sigma_x 
	+ a_y \sigma_y 
	+ a_z \sigma_z )
	+ \frac{1}{2} (1-p) 
	( I 
	- a_x \sigma_x 
	- a_y \sigma_y 
	+ a_z \sigma_z )
	\nonumber \\
&=&	
	\begin{pmatrix}
		\rho_{00} & ( 2p - 1) \rho_{01} \\
		( 2p - 1) \rho_{10} & \rho_{11} 
	\end{pmatrix}.
	\label{flip matrix form}
\end{eqnarray}
The first expression in the right hand side shows that
the phase is left unchanged with probability $p$ while
it is flipped as
$ e^{i \phi} \to e^{i(\phi + \pi)} = - e^{i \phi } $
with probability $ 1-p $.
Hence it is natural to call this a phase flip channel.
In particular, when $ p = \frac{1}{2} $,
the transverse components $ ( a_x, a_y ) $ of the Bloch vector,
or the off-diagonal elements $ \rho_{01} =\rho_{10}^*$,
vanish after the channel is applied
and the information about the phase is completely lost.
However, 
the diagonal elements $ \rho_{00} $ and $ \rho_{11} $,
which represent populations of spins in the states 
$ | 0 \ket $ and $ | 1 \ket $, respectively,
remain unchanged.
Due to these properties,
decoherence generated via the phase flip channel 
is called phase decoherence.

It should be noted that
different sets of an initial state of the environment
and a unitary operator of the whole system 
may yield the same channel.
Instead of Eq.~(\ref{psi_e}) and (\ref{U for flip}),
we may take a mixed environment state
\begin{equation}
	\rho_{\rm e} = p |+\rangle \langle +| + (1-p) |- \rangle 
	\langle -| = \frac{1}{2} 
	\begin{pmatrix}
	1&2p-1\\
	2p-1&1
	\end{pmatrix},
\end{equation}
where $|+ \rangle$ and $|- \rangle$ are the normalized eigenvectors of
$\sigma_x$ with eigenvalues $1$ and $-1$ respectively,
and the controlled {\sc not} gate
\begin{equation}
	U_{\mbox{\tiny CNOT}} 
	= 
	\begin{pmatrix}
	1 & 0 & 0 & 0 \\
	0 & 1 & 0 & 0 \\
	0 & 0 & 0 & 1 \\
	0 & 0 & 1 & 0 
	\end{pmatrix} 
	= I \otimes | + \rangle \langle +| 
	+ \sigma_z \otimes |- \rangle \langle -|.
\end{equation}
Substituting them into Eq.~(\ref{channel}) we obtain again
the phase flip channel Eq.~(\ref{flip matrix form}).

\subsection{Random fluctuating field}
\label{sec:rff}

It is possible to reproduce the phase flip channel
without resort to the partial trace.
This is done by introducing the mixing process defined previously.

Let us consider a single spin Hamiltonian 
\begin{eqnarray}
	\label{rf}
	H_{\rm rf} =  - \omega(t) \frac{\sigma_z}{2},
	\label{Hrf}
\end{eqnarray}
where $ \omega(t) $ is a randomly fluctuating field.
We have taken the natural unit $ \hbar = 1 $. 
The time-evolution operator associated with the Hamiltonian (\ref{Hrf}) is
the phase shift gate
% Let us define the phase shift gate
\begin{equation}
	S( \theta ) = e^{ i \theta \sigma_z /2}
\end{equation}
with
\begin{eqnarray}
	\theta = \int_0^t \omega (\tau) \, d \tau .
\end{eqnarray}
The phase $\theta$ integrates 
the effect of $\omega( \tau )$ in the interval $[0, t]$.
In the context of NMR,
the phase shift gate is implemented with a longitudinal 
magnetic field or a scalar coupling with another spins as we 
discuss later.
The phase shift gate acts on the density matrix as
\begin{equation}
	S( \theta ) \rho_{\rm s} \, S^\dagger ( \theta )
	=
	\begin{pmatrix}
		\rho_{00} & e^{i \theta} \rho_{01} \\
		e^{-i \theta} \rho_{10}  & \rho_{11} 
	\end{pmatrix}.
\end{equation}
Given a probability distribution $ p( \theta ) $ which
characterizes the random fluctuating field,
the mixing process is evaluated as~\cite{Leung99}
\begin{equation}
	{\mathscr M} ( \rho_{\rm s} )
	=
	\int_{- \infty}^\infty \!\!
	p ( \theta )
	S( \theta ) \rho_{\rm s} \, S^\dagger ( \theta )
	d \theta
	=
	\begin{pmatrix}
		\rho_{00} & \bra e^{i \theta} \ket \rho_{01} \\
		\bra e^{-i \theta} \ket \rho_{10}  & \rho_{11} 
	\end{pmatrix}.
	\label{mixing with phase shift}
\end{equation}
For any probability distribution $ p(\theta) $,
$$
	\Big| \bra e^{-i \theta} \ket \Big|
	=
	\left|
		\int_{- \infty}^\infty \!\!
		p ( \theta ) \,  e^{-i \theta} d \theta
	\right|
	\le
	\int_{- \infty}^\infty \!
	\Big| p ( \theta ) \,  e^{-i \theta} \Big|  d \theta
	= 1.
$$
Therefore the absolute value of the off-diagonal elements of
$ {\mathscr M} ( \rho_{\rm s} ) $
become smaller than those of $  \rho_{\rm s} $.
When the average $ \bra e^{-i \theta} \ket $ is a real number, 
the map
$ {\mathscr M} ( \rho_{\rm s} ) $ reproduces 
the phase flip channel (\ref{flip matrix form}). This applies
when $p(\theta) = p(-\theta)$ for example.

We can further simplify the model without losing the essence of the
random fluctuating field model. Suppose that
$\omega(t)$ takes only two values, $\omega_0 \pm \delta\omega$,
with the corresponding probabilities $p(\pm \delta\omega)$. 
We simulate phase decoherence phenomena 
according to this simplified random fluctuating field model. 

Relaxation phenomena for nuclear spin qubits, including phase decoherence,
have long been studied in somewhat different context 
in non-equilibrium statistical physics and employed as probes 
in condensed matter physics~\cite{ref:rff}.
Studies on a relaxation process induced by environmental fluctuations 
have played an essential role when we want to extract information 
on the environments. Conversely, we can evaluate the
time evolution of a system provided that the structure of 
the environment as well as the interaction between the system and 
the environment are given. Spin relaxation has often been analyzed 
by the phenomenological Bloch equations.
The equations are characterized by two parameters 
called the longitudinal relaxation time $T_1$ and 
the transverse relaxation time $T_2$. 
The former characterizes energy relaxation 
whereas the latter describes phase decoherence.
Relaxation phenomena can also be described by the method of master 
equation. This is called a Redfield equation in the field of spin relaxation. 
Both the Bloch and the Redfield equations are valid only for 
the so-called 
narrowing limit where the characteristic time $\tau_{\rm e}$ of the environment
is very short resulting in an exponential decay in the relevant 
spin variables. 
In contrast, the random fluctuating field model,
though phenomenological, is applicable to any 
time scale 
ranging from the narrowing limit ($\tau_{\rm e} \rightarrow 0$) to the 
slow modulation with large $\tau_{\rm e}$.
Therefore, it is reasonable to take the random fluctuating field model
as a basis for simulating phase decoherence phenomena. 

\subsection{Suppressing decoherence by the bang-bang control }

Several groups~\cite{bb,uchiyama} have proposed and analyzed
a useful technique to suppress decoherence,
which is called a quantum bang-bang control.
We briefly explain the principle of the bang-bang control.
When a qubit system evolves in time,
the interaction with its environment usually causes decoherence 
in the qubit state. 
If, however, time-evolution of the qubit could be reversed 
by some methods, the qubit returns to its initial state 
and decoherence would be eliminated. Concerning the phase 
decoherence, a time-reversal operation can be simply 
implemented with a pair of $ \pi $-pulses assuming 
that the state of environment remains unchanged between two pulses. 
The $ \pi $-pulses around the $ x $- and $ -x $-axes transform
the qubit state with unitary operators
\begin{equation}
	V         = e^{-i \pi \sigma_x / 2},
	\qquad
	V^\dagger = e^{ i \pi \sigma_x / 2},
	\label{pi pulse}
\end{equation}
respectively.
The phase shift operator $ S(\theta) = e^{ i \theta \sigma_z /2} $
has the property
\begin{equation}
	V^\dagger S(\theta) V = S(-\theta) = S(\theta)^{-1}.
	\label{property of V}
\end{equation}
Hence, by inserting a pair of $ \pi $-pulses 
in a product of phase shift operators $ S(\theta) $
we get
\begin{equation}
	V^\dagger S(\theta) V S(\theta) = I.
\end{equation}
Therefore, the state $ \rho_{\rm s} $ of the qubit comes back 
to the initial one as
\begin{equation}
	\big( V^\dagger S(\theta) V S(\theta) \big)
	\rho_{\rm s}
	\big( V^\dagger S(\theta) V S(\theta) \big)^\dagger
	= \rho_{\rm s}.
\end{equation}
The phase shift is also canceled
if a pair of $ \pi $-pulses is inserted as
\begin{equation}
	S( \theta_2) V^\dagger S(\theta_2 + \theta_1) V S(\theta_1) 
	= I
\end{equation}
showing that
the locations of $ \pi $-pulse insertions may be chosen rather arbitrarily.

The time-evolution operator $ S( \omega t ) $ affects the qubit state
if the interaction with environment causes
a phase shift proportional to time.
Here $ \omega $ is a parameter which characterizes 
the environment state
and strength of interaction between the system and the environment.
Let us introduce a time interval $ t_{\rm b}$
and put $ 2n t_{\rm b} = t $ 
with a positive integer $ n $.
Then we have
\begin{equation}
	\underbrace{
		S( \omega t_{\rm b} ) \cdot
		S( \omega t_{\rm b} ) \cdots
		S( \omega t_{\rm b} ) \cdot
		S( \omega t_{\rm b} ) 
	}_{2n}
	= S( \omega t ).
\end{equation}
By inserting $ \pi $-pulses we recover the initial state since
\begin{equation}
	V^\dagger S( \omega t_{\rm b} )
	V S( \omega t_{\rm b} ) 
	\cdots
	V^\dagger S( \omega t_{\rm b} )
	V S( \omega t_{\rm b} ) 
	= I.
\end{equation}
This argument indicates that phase decoherence is suppressed
by applying a regular sequence of $ \pi $-pulses on the qubit.

In a general circumstance, 
the environment state is not stationary
and the phase shift is not proportional to time.
Then $ S( \omega t ) $ is replaced with
\begin{equation}
	U(t;t_0)
	= S \left( \int_{t_0}^t \omega (\tau) d \tau \right).
	\label{U(t;t_0)}
\end{equation}
Even in such a general case,
if $ \omega(t) $ dose not vary rapidly
and remains almost constant during the short time interval $ t_{\rm b} $,
it is legitimate to use approximation
\begin{equation}
	U(t_0 + 2t_{\rm b}; t_0 + t_{\rm b}) 
	\approx
	U(t_0 + t_{\rm b}; t_0) 
\end{equation}
and hence
\begin{equation}
	V^\dagger 
	U(t_0 + 2t_{\rm b}; t_0 + t_{\rm b}) 
	V 
	U(t_0 + t_{\rm b}; t_0) 
	\approx
	I.
\end{equation}
Therefore the phase shift will be mostly canceled by inserting $ \pi $-pulses 
with a short interval $ t_{\rm b} $ in the time-evolution operator
(\ref{U(t;t_0)})
and hence the associated decoherence will be suppressed.

%=====================================================================
\section{Two-qubit system as a composite of system and environment}
%=====================================================================
\label{sec:two-qubit}

\subsection{Artificial environment}

\begin{figure}[b]
\begin{center}
\includegraphics[bb=80 40 520 290,width=7cm]{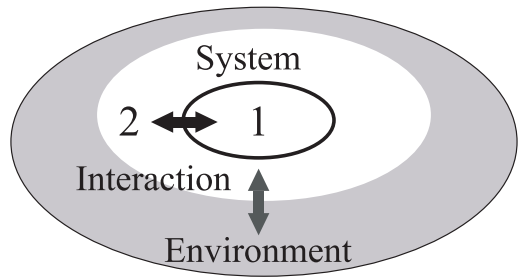}
\end{center}
\caption{
\label{fig1}
System and its environment. 
The system consists of two subsystems,~1 and~2.
}
\end{figure}

Any system in an environment has a Hamiltonian of the form
\begin{equation}
	H_{\rm t} = H_{\rm s} + H_{\rm e}+ H_{\rm se},
\end{equation}
where $ H_{\rm s} $ and $ H_{\rm e} $
govern intrinsic behaviors 
of the system and the environment, respectively, while
$ H_{\rm se} $ represents interaction between them.
% Decomposition of the Hamiltonian in this manner
The relation of the system and the environment 
is schematically depicted in Fig.~\ref{fig1}. 
Zurek~\cite{zurek}, for example, discussed 
a simple model where 
a one-qubit system is coupled to an $ n $-qubit environment
through interaction of the form $ \sigma_z \otimes \sigma_z $.

Suppose the interaction $ H_{\rm se} $ is so weak 
that its effect on the system qubit is negligible compared 
with that of $ H_{\rm s} $ for a certain time scale $ \tau $.
Assume further that 
the system consists of two subsystems, 
which are referred to as subsystems~1 and~2.
Then the system Hamiltonian $ H_{\rm s} $ is decomposed as
\begin{equation}
	H_{\rm s} = H_{1} + H_{2} + H_{12}.
\end{equation}
Here $ H_{1} $ and $ H_{2} $ govern
intrinsic behaviors of the subsystems,
while $ H_{12} $ describes interaction between them.
Under this decomposition, 
we may regard
the subsystem~1 as a new system and
the subsystem~2 as an artificial environment.
The subsystem~1 will exhibit a decoherence-like behavior
if the subsystem~2 simulates an environment
that has many degrees of freedom.

Particularly, nuclear spins used in NMR
have long relaxation times of the order of $ \tau \sim 10 $~s.
Thus nuclear spins are almost isolated from the environment 
in the time scale $ \tau $.
In such a circumstance
it is legitimate to regard 
some of the spins as an artificial environment for the other spins.

Zhang {\it et al.}~\cite{zhang} 
experimentally studied the spin dynamics of 
${}^{13}$C-labeled trichloroethane,
which has three spins in a molecule.
They regarded three spins 
as a composite of a two-qubit system and a one-qubit environment.
They claimed that they observed decoherence in the two-qubit system. 
However, the artificial environment must have
a large number of degrees of freedom 
to introduce irreversible decoherence-like behavior in the system while
the one-qubit environment employed in their experiments had not enough
degrees of freedom.
Hence their system exhibited a periodic behavior
and failed to introduce an irreversible change in the system.
Ryan {\it et al.}~\cite{ryan}
overcame this difficulty by employing 
a seven-qubit molecule for simulating complex dynamics of an environment.

Teklemariam {\it et al.}~\cite{Teklemariam}
proposed to apply a stochastic classical field~\cite{ernst} 
on the artificial environment 
to generate artificial decoherence.
Although the artificial environment has 
only a few degrees of freedom,
it can simulate an open environment 
if it is randomized by an external stochastic field.
With this idea, they observed decoherence-like behavior by using NMR.
Although we closely follow Teklemariam {\it et al.},
we simplify 
the principle for generating artificial phase decoherence phenomena 
to the limit. Thanks to this simplification, we are able to
analyze the phenomena without numerical calculations 
and compare theoretical and experimental results directly.
% Moreover, we test the bang-bang control for suppressing decoherence.

Several remarks are in order.
An experimental work that may be called 
demonstration of ``quantum bang-bang control'' was only 
reported by Morton  {\it et al.}~\cite{Morton} to the best of our knowledge, 
where the time development of Rabi oscillation 
was undone by successive $\pi$-pulses. 
Their experiments are closely related with 
the ``quantum Zeno effect'' experiment
by Itano  {\it et al.}~\cite{Itano}, 
which was in turn based on the proposal by Cook~\cite{Cook}. 
Cook pointed out that demonstration 
of ``quantum Zeno effect'' is difficult and thus proposed 
an experiment inhibiting Rabi oscillation, 
not relaxation, by frequent measurements.
 
We emphasize that our experiments really demonstrate 
the suppression of relaxation, albeit artificial, 
and are not mere suspension of Rabi oscillation \cite{Morton,Itano}.   
Although our techniques employed in this work are based on
the well-known spin-decoupling technique in NMR~\cite{ernst},
we propose the new usage of this technique 
in understanding relaxation phenomena experimentally
and in providing a test bench to develop indispensable techniques
in quantum information processing.       

We list preceding experimental works dealing with engineered noise here. 
Kohmoto {\it et al.}~\cite{Ref-FK} analyzed  spin relaxation induced 
by experimentally generated classical random field.
Viola {\it et al.}~\cite{NS} demonstrated the noiseless subsystem 
with NMR, in which collective noise was engineered 
through gradient-diffusion method. 
Kwiat {\it et al.}~\cite{DFS} employed the collective artificial noise
in their decoherence-free subspace experiments with linear 
optics. 
Kielpinski {\it et al.}~\cite{ion} applied collective noise to ions 
by irradiating laser light on the ions.
Since our noise is generated through interaction 
between qubits, our experiment is closer to realistic situation 
than theirs.

\subsection{Two-qubit system}

In the rest of this paper we shall study a two-qubit system.
Each qubit is referred to as qubit~1 and qubit~2, respectively.
Qubit~1 is regarded as a system while qubit~2 
as an environment coupled to qubit~1. 
We use the Hamiltonian
\begin{equation} 
	H =
	J I_z \otimes I_z,
	\label{tilde H}
\end{equation}
where $ I_k = \sigma_k / 2 $ $ (k=x,y,z) $ and $J$ specifies 
the strength of the interaction between the two qubits. 
We assume $J > 0$ without loss of generality.  
The Hamiltonian
of a two-spin molecule in a proper rotating frame has 
this form as will be discussed in Section~\ref{rf_ham}. 

When the Hamiltonian (\ref{tilde H}) acts on states
$ | \psi \ket \otimes | 0 \ket $ and
$ | \psi \ket \otimes | 1 \ket $, it yields
\begin{eqnarray}
\begin{array}{ccrc}
	H | \psi \ket \otimes | 0 \ket &
	=&
	  \frac{J}{2} I_z | \psi \ket \!\! & \!\! \otimes \, | 0 \ket,
	\label{H|0>}
	\\
	H | \psi \ket \otimes | 1 \ket &
	=&
	- \frac{J}{2} I_z | \psi \ket \!\! & \!\! \otimes \, | 1 \ket,
	\label{H|1>}
\end{array}
\label{H|>}
\end{eqnarray}
respectively.
Thus the Hamiltonian $ H $ describes 
an effective magnetic field acting on qubit~1.
The effective magnetic field is $J/2$
when qubit~2 is in the state $ | 1 \ket $
and it is $-J/2$ when qubit~2 is in $ | 0 \ket $.
Hence, by flipping qubit~2 randomly,
we can realize 
a random fluctuating field for qubit~1, see \S~\ref{sec:rff}. 
We will extensively use
this fact in the following.

In the next two subsections, we show two examples of 
artificially generated phase decoherence phenomena: 
One is phase decoherence of a single qubit in a transmission line, 
and the other is that in a quantum memory. 
The difference between these two examples 
is that of the characteristic of the random fields. We emphasize that 
any phase decoherence phenomena can be generated by 
controlling qubit~2.

\subsection{Phase decoherence in a quantum transmission line
and its suppression by the bang-bang control}
Let us imagine a situation in which
a flying qubit passes through a quantum transmission line
where noise acts on the qubit as it propagates.
The noise source is assumed to localize at a certain
region in the line \cite{kitajima}.

We construct a model which realizes the above situation
with the two-qubit system discussed in the previous subsection.
We regard qubit~1 as a flying qubit and qubit~2 as an environment.
Suppose that the initial state of qubit~2 is $ | 0 \ket $.
Qubit~2 is flipped to $ | 1 \ket $ at time $ t_1 $, 
which corresponds to a position in the transmission line
where noise is switched on, 
and it is flipped back to $ | 0 \ket $ at time $ t_1 + \Delta $, 
which corresponds to a position
where noise is switched off.
We then observe qubit~1 at later time $ T $ $ ( > t_1 + \Delta ) $.
The state of qubit~1 at $T$ is determined by applying
the phase shift
\begin{equation}
\label{eq:ps_T}
       e^{ -i \frac{J}{2} I_z (T-t_1-\Delta)} e^{ i \frac{J}{2} I_z \Delta}
       e^{ -i \frac{J}{2} I_z t_1}
	= S ( J \Delta ) S \left(- \frac{J}{2} T \right)
\end{equation}
on the initial state.
Now we regard the time interval $ \Delta $ as a random variable.
Assume that $ \Delta $ takes its value
in the range $ 0 \le J \Delta \le 2 \pi $ 
with uniform probability distribution.
Then the mixing process (\ref{mixing with phase shift}) yields
\begin{equation}
	{\mathscr M} ( \rho_{\rm s} )
	=
	\frac{1}{2 \pi}
	\int_{0}^{ 2 \pi } \!\!
	S( \theta ) \rho_{\rm s} \, S^\dagger ( \theta )
	d \theta
	=
	\begin{pmatrix}
		\rho_{00} & 0 \\
		0 & \rho_{11} 
	\end{pmatrix}.
	\label{decoherence by artificial channel}
\end{equation}
Thus, a complete decoherence takes place in qubit~1.

Let us apply the bang-bang control to qubit~1 
in the artificial phase flip channel
introduced above.
Assume that the initial state of qubit~2 is $ | 0 \ket $.
Apply $ \pi $-pulse sequence,
$ V $ and $ V^\dagger $ of (\ref{pi pulse}), repeatedly
with constant time interval $ t_{\rm b} $ on qubit~1
during $ 0 \le t \le T $.
Qubit~2 turns to $ | 1 \ket $ at time $ t_1 $
and turns back to $ | 0 \ket $ at time $ t_1 + \Delta $.
The time $ t_1 $ and the interval $ \Delta $ are random variables.
The value of $ \Delta $ varies in
the range 
$ 0 \le J \Delta \le 2 \pi $ 
with uniform probability distribution.
Measure the state of qubit~1 at time $ T $ and
take an average of the data with respect to
the random variable $ \Delta $.
What is the resulting state $ {\mathscr M} ( \rho_{\rm s} ) $
for this mixing process?

We consider four cases separately:\\
(1)~there are even number of $ \pi $-pulses in $ [0,  t_1] $
and even number of $ \pi $-pulses in $ [t_1, t_1 + \Delta] $;\\
(2)~even in $[ 0,  t_1] $
and odd in $  [t_1, t_1 + \Delta] $;\\
(3)~odd in $ [0,  t_1] $
and even in $  [t_1, t_1 + \Delta] $;\\
(4)~odd in $[0,  t_1]  $
and odd in $  [t_1, t_1 + \Delta] $.\\
Define $ \varepsilon_0 $ such that the first $ \pi $-pulse
in $ t_1 \le t \le t_1 + \Delta $ is applied at $ t = t_1 + \varepsilon_0 $.
Similarly, 
define $ \varepsilon_1 $ such that the final $ \pi $-pulse
in $ t_1 \le t \le t_1 + \Delta $ is applied at 
$ t = t_1 + \Delta - \varepsilon_1 $.
If the number of $ \pi $-pulses in $ t_1 \le t \le t_1 + \Delta $ is $ m $,
\begin{equation}
	\Delta =
	\varepsilon_0 + (m-1) t_{\rm b} + \varepsilon_1.
\end{equation}
Assume that $ J t_{\rm b} \ll 2 \pi $
so that there are sufficiently many pulses in the interval $[t_1,
t_1+ \Delta ]$. Under this assumption,
the variables $ \varepsilon_0 $ and $ \varepsilon_1 $ are regarded as
random variables
taking values in the range $ 0 \le \varepsilon_i \le t_{\rm b} $
with uniform probability distribution.

\begin{figure}[t]
\begin{center}
\includegraphics[width=8cm]{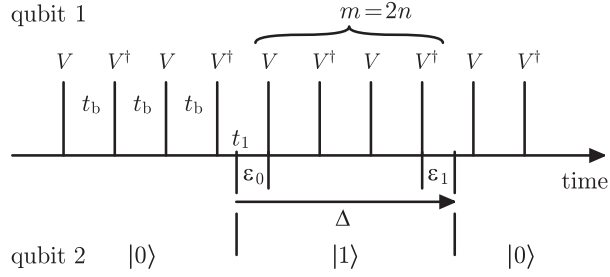}
\end{center}
\caption{
\label{fig:pipi}
Bang-bang pulses, $ V $ and $ V^\dagger $,
are applied repeatedly on qubit~1.
Qubit~2 is flipped to $ | 1 \ket $ at $ t_1 $
and is flipped back to $ | 0 \ket $ at $ t_1 + \Delta $.
This depicts $ U_1 $ of Eq.~(\ref{U_1}) in the case~(1).
}
\end{figure}

The time-evolution operator for qubit~1 is calculated
for each case as follows.
If we put 
\begin{equation}
	 L( \tau ) 
	= S \left( \frac{J}{2} \tau \right)
	= e^{i \frac{J}{2} I_z \tau},
	\label{L}
\end{equation}
the time-evolution is generated by (\ref{H|>}) as
\begin{eqnarray}
\begin{array}{ccrc}
	e^{-iH \tau} | \psi \ket \otimes | 0 \ket &
	=&
	L( - \tau ) | \psi \ket \!\! & \!\! \otimes \, | 0 \ket,
	\\
	e^{-iH \tau} | \psi \ket \otimes | 1 \ket &
	=&
	L(   \tau ) | \psi \ket \!\! & \!\! \otimes \, | 1 \ket.
\end{array}
\label{e^H = L}
\end{eqnarray}
The sequence of $ \pi $-pulse pairs is represented by 
alternate insertions of $ V $ and $ V^\dagger $ of Eq.~(\ref{pi pulse})
in the time-evolution operator product.
For the case~(1), the number of $ \pi $-pulses is $ m = 2n $.
The time-evolution operator is calculated 
with a help of Fig.~\ref{fig:pipi} as
\begin{eqnarray}
	U_{1}
&=&
	V^\dagger L(-t_{\rm b}) V
	L(-(t_{\rm b}-\varepsilon_1))
      L(\varepsilon_1) V^\dagger L(t_{\rm b}) 
 	\Big[ L(t_{\rm b}) V^\dagger L(t_{\rm b}) V \Big]^{n-1}
	L(\varepsilon_0)L(-(t_{\rm b}-\varepsilon_0))
	\nonumber \\
&=&
	L( t_{\rm b} -(t_{\rm b}-\varepsilon_1)+
        \varepsilon_1 -t_{\rm b}
        + \varepsilon_0-(t_{\rm b}-\varepsilon_0))
	\nonumber \\
&=&
	L(2 \varepsilon_1  + 2 \varepsilon_0 - 2 t_{\rm b}),
	\label{U_1}
\end{eqnarray}
where use has been made of the property of $ V $, Eq.~(\ref{property of V}).
For the case~(2) with $ m = 2n + 1 $, we obtain
\begin{eqnarray}
	U_{2}
&=&
	V^\dagger L(-(t_{\rm b}-\varepsilon_1))
        L(\varepsilon_1) V
	\Big[ L(t_{\rm b}) V^\dagger L(t_{\rm b}) V \Big]^{n}
	L(\varepsilon_0)L(-(t_{\rm b}-\varepsilon_0))
        \nonumber \\
&=&
	L(-2 \varepsilon_1 + 2 \varepsilon_0).
	\label{U_2}
\end{eqnarray}
For the case~(3) with $ m = 2n $,
\begin{eqnarray}
	U_{3}
&=&
	V^\dagger L(-(t_{\rm b}-\varepsilon_1))
        L(\varepsilon_1) V L(t_{\rm b}) V^\dagger 
	\Big[ L(t_{\rm b}) V L(t_{\rm b}) V^\dagger \Big]^{n-1}
	L(\varepsilon_0)L(-(t_{\rm b}-\varepsilon_0))V L(-t_{\rm b}) 
        \nonumber \\
&=&
        L(-2 \varepsilon_1 - 2 \varepsilon_0 + 2 t_{\rm b}).
\end{eqnarray}
Finally for the case~(4) with $ m = 2n + 1 $,
\begin{eqnarray}
	U_{4}
&=&
	V^\dagger L(-t_{\rm b}) V L(-(t_{\rm b}-\varepsilon_1))
        L(\varepsilon_1)V^\dagger  
	\Big[ L(t_{\rm b}) V L(t_{\rm b}) V^\dagger  \Big]^{n}
	L(\varepsilon_0)L(-(t_{\rm b}-\varepsilon_0)) V L(-t_{\rm b}) 
        \nonumber \\
&=&
	L(2 \varepsilon_1 - 2 \varepsilon_0).
\end{eqnarray}
By taking average with respect to $ \varepsilon_0 $ and 
$ \varepsilon_1 $, and also average over the four cases,
the mixing process (\ref{mixing}) yields
\begin{eqnarray}
	{\mathscr M} ( \rho_{\rm s} )
	\!\! &=& \!\!
	\frac{1}{4}
	\frac{1}{  t_{\rm b}^2}
	\int_{0}^{ t_{\rm b} } \!\! d \varepsilon_0
	\int_{0}^{ t_{\rm b} } \!\! d \varepsilon_1
	\Big(
		  U_1 \rho_{\rm s} U_1^\dagger
		+ U_2 \rho_{\rm s} U_2^\dagger
		+ U_3 \rho_{\rm s} U_3^\dagger
		+ U_4 \rho_{\rm s} U_4^\dagger
	\Big).
	\qquad
\end{eqnarray}
Each term in the parentheses is calculated as;
\begin{eqnarray}
	L( 2 \varepsilon ) \rho_{\rm s} L^\dagger ( 2 \varepsilon )
&=&
	S( J \varepsilon ) \rho_{\rm s} S^\dagger ( J \varepsilon )
	\nonumber \\
&=&
	\begin{pmatrix}
		\rho_{00} & e^{ i J \varepsilon} \rho_{01} \\
		e^{-i J \varepsilon} \rho_{10}  & \rho_{11} 
	\end{pmatrix}
	\label{L rho L}
\end{eqnarray}
while the integral is evaluated as
\begin{equation}
	\frac{1}{t_{\rm b}} 
	\int_{0}^{ t_{\rm b} } \!\! d \varepsilon
	\, e^{ i J \varepsilon}
	=
	\frac{ e^{ i J t_{\rm b}} - 1 }{i J t_{\rm b}}
	=
	\frac{\sin (J t_{\rm b}/2)}{ J t_{\rm b}/2 } \,
	e^{ i J t_{\rm b} /2}.
\end{equation}
By combining these results we finally obtain
\begin{eqnarray}
	{\mathscr M} ( \rho_{\rm s} )
	&=&
	\begin{pmatrix}
		\rho_{00} & \kappa \rho_{01} \\
		\kappa \rho_{10}  & \rho_{11} 
	\end{pmatrix},
	\label{suppressed by bang-bang}
	\\ 
        \kappa 
	&=& 
	\left\{ 
		\frac{ \sin (J t_{\rm b}/2)}{ J t_{\rm b}/2 } 
	\right\}^{\! 2}.
	\label{kappa}
\end{eqnarray}
Since we have already assumed that $J t_{\rm b} \ll 2\pi$, 
$\kappa $ should be approximately unity. 
Comparing this result
(\ref{suppressed by bang-bang})
with
(\ref{decoherence by artificial channel}),
we see that the bang-bang control suppresses phase decoherence.
%In particular, by making the interval of the bang-bang pulses 
%infinitesimally short,
%$ t_{\rm b} \to 0 $,
%we get $ \kappa \to 1 $, 
%so the decoherence is totally eliminated and
%the initial state is completely preserved.

\subsection{Phase decoherence in a quantum memory
and its suppression by the bang-bang control}

Suppose a qubit sits in a quantum memory device (quantum register).
The qubit is exposed to a noisy environment and loses its phase coherence.
The noise is described by the random fluctuating field 
in the Hamiltonian~(\ref{rf}).
We assume that the field variable $ \omega (t) $ in~(\ref{rf}) 
has a white noise spectrum.
This can also be interpreted
as a model of the phase relaxation process in NMR~\cite{random_walk}.
The phase $\theta$ of the qubit evolves 
in time and is randomly distributed at a later time. 
The distribution function $p(\theta)$ of $\theta$ is Gaussian,  
\begin{eqnarray}  
\label{rw_d}
p(\theta) = \frac{1}{\sqrt{2\pi}s}e^{-\theta^2/2s^2},
\end{eqnarray}
where $ s^2 $ is proportional to the evolution time $t$~\cite{random_walk}. 

We can also construct a model which realizes the above situation
with the previously introduced
two-qubit system by modifying the random field properly.
We regard qubit~1 as a qubit in a register
and qubit~2 as an environment.
Set the initial state of qubit~2 to $ | 0 \ket $ at time $ t_0 = 0 $,
turn it to $ | 1 \ket $ at $ t_1 $,
turn it back to $ | 0 \ket $ at $ t_2 $,
and repeat flipping qubit~2
at $ t_3 $, $ t_4 $ and so on.
Under this manipulation qubit~2 effectively works as a noisy environment 
acting on
qubit~1. The time interval between consecutive flippings is denoted as
\begin{equation}
	\Delta_j
	= t_{j+1} - t_j
	= \bar{\Delta} ( 1 + \alpha \xi_j )
	\quad
	( j = 0, 1, 2, \cdots ).
	\label{random interval}
\end{equation}
Here $ \{ \xi_j \} $ are independent random variables obeying the
probability distribution function
\begin{equation}
	p ( \xi_j )
	= \frac{1}{ \sqrt{2 \pi} } \, e^{- \xi_j^2 / 2 }
	\label{Gaussian}
\end{equation}
in parallel with (\ref{rw_d}). $ \bar{\Delta} $ is 
the average of $ \{ \Delta_j \} $.
The parameter $ \alpha $ $ ( 0 \le \alpha \le 1/4 ) $ 
characterizes variance of the time intervals.
To ensure that $ \Delta_j $ in (\ref{random interval}) is positive,
the range of $ \xi_j $ should be $ - \frac{1}{\alpha} < \xi_j $.
However, if $ \alpha $ is not too large,
the probability of having negative $ \Delta_j $ is negligibly small.
Hence, it is legitimate 
to extend the range of $ \xi_j $-integration to $ - \infty < \xi_j < \infty $
when we take an average.
At time $ t_{2n} $ the evolution operator for qubit~1 becomes
\begin{eqnarray}
	U_{\xi}
&=&	L ( -\Delta_1      + \Delta_2 + \cdots 
            -\Delta_{2n-1} + \Delta_{2n}).
	\quad
	\label{U_xi}
\end{eqnarray}
In this case, the mixing process (\ref{mixing}) yields
\begin{eqnarray}
	{\mathscr M} ( \rho_{\rm s} )
&=&
	\int_{- \infty}^{\infty} \!\! d \xi_1
	\int_{- \infty}^{\infty} \!\! d \xi_2
	\cdots
	\int_{- \infty}^{\infty} \!\! d \xi_{2n} \quad
	p( \xi_1 )
	p( \xi_2 ) \cdots
	p( \xi_{2n} ) \:
	U_{\xi} \rho_{\rm s} U_{\xi}^\dagger.
	\label{it}
\end{eqnarray}
The matrix
$ U_{\xi} \rho_{\rm s} U_{\xi}^\dagger $
is calculated similarly to Eq.~(\ref{L rho L}).
The integral in the off-diagonal components is evaluated as
\begin{eqnarray}
\lambda
	&=&
	\int_{- \infty}^\infty \! d \xi_1 \, p ( \xi_1 ) \,
	\int_{- \infty}^\infty \! d \xi_2 \, p ( \xi_2 ) \,
	   e^{-i (J/2) \bar{\Delta} ( 1 + \alpha \xi_1 ) }
	   e^{ i (J/2) \bar{\Delta} ( 1 + \alpha \xi_2 ) }  
	=
	\, e^{ - \frac{1}{4} ( J \bar{\Delta} \alpha )^2 }.
\end{eqnarray}
Thus (\ref{it}) becomes
\begin{eqnarray}
&&	{\mathscr M} ( \rho_{\rm s} )
	=
	\begin{pmatrix}
		\rho_{00} 
		& 
		\lambda^n \rho_{01} 
	\\
		\lambda^n \rho_{10}
		& 
		\rho_{11} 
	\end{pmatrix},
\end{eqnarray}
at $ \overline{ t_{2n} } = 2n \bar{\Delta} $, 
the average of time $ t_{2n} $. 
Then the absolute value of the matrix element $ \rho_{01} $ 
in $ {\mathscr M} ( \rho_{\rm s} ) $ is multiplied by
\begin{eqnarray}
	\lambda^n
&=&
	\exp
	\bigg\{ \!\!
		- \frac{1}{4} 
		( J \bar{\Delta} \alpha )^2 
		\frac{\overline{t_{2n}}}{2 \bar{\Delta}}
	\bigg\}
	\nonumber \\
&=&
	\exp
	\Big( \!
		- \frac{1}{8} 
		J^2 \alpha^2 \bar{\Delta} \, \overline{t_{2n}}
	\Big)
	= e^{-\overline{t_{2n}} / T_2^*}.
	\label{prediction of random fluc}
\end{eqnarray}
The last line defines  $ T_2^* $
which we call the effective transverse relaxation time
since $ {\mathscr M} ( \rho_{\rm s} ) $ 
is calculated only at $\overline{ t_{2n}}$. 
We note that $\bar{\Delta}$ is regarded as the correlation time 
of the artificial environment and that the decay of $|\rho_{01}|$ 
is non-exponential for $t \sim \bar{\Delta}$.
Thus, we see that phase decoherence in the presence of
random fluctuating field
is characterized by the decay constant
\begin{equation}
	T_2^*
	=
	\frac{8}{ J^2 \alpha^2 \bar{\Delta} }.
	\label{prediction of T2}
\end{equation}
It is clear that
$ T_2^* $ becomes smaller for the larger variance $ \alpha $
of fluctuation of the effective magnetic field.

Now let us apply the bang-bang control along
with the artificial random fluctuating field.
Assume that the bang-bang pulse interval $ t_{\rm b} $
is short enough so that $ t_{\rm b} \ll \alpha \bar{\Delta} $
is satisfied.
Then the argument of the previous subsection to evaluate
$ {\mathscr M} ( \rho_{\rm s} ) $ is applicable here as well. 
For $ \overline{t_{2n}} = 2n \bar{\Delta} $,
there are $ n $ random switching of qubit~2 
from $|0\ket$ to $|1\ket$ and also $n$ random switching from
$|1 \rangle$ back to $|0 \rangle$ on average.
After one cycle of switching of qubit~2, 
the element $ \rho_{01} $ of the density matrix
is multiplied by the factor $ \kappa $ given by Eq.~(\ref{kappa}).
Therefore the off-diagonal element $ \rho_{01} $ is multiplied by
\begin{equation}
\label{def_t2b}
	\kappa^n
	= 
	\kappa^{ \overline{t_{2n}} / ( 2 \bar{\Delta}) }
	=
	e^{ - \overline{t_{2n}} / T_{2 \rm b}^* }
\end{equation}
at $ \overline{t_{2n}} $. 
Equation~(\ref{def_t2b}) defines $ T_{2 \rm b}^* $, 
the decay constant that characterizes the phase decoherence
under the bang-bang control. It is explicitly given as
\begin{equation}
	T_{2 \rm b}^*
	=
	-\frac{2 \bar{\Delta}}{ \ln \kappa}
	=
	\bar{\Delta}
	\left\{ 
		\ln 
		\Big| 
		\frac{ J t_{\rm b}/2 }{ \sin (J t_{\rm b}/2) }
		\Big|
	\right\}^{-1}.
	\label{bang-bang prediction}
\end{equation}
Since $ \kappa $ approaches 1 from below
when $t_{\rm b} \to 0$,
$ T_{2 \rm b}^* $ approaches $\infty$ in this limit.
Therefore decoherence is suppressed  
by bang-bang pulses.

Here we would like to give a remark on
the work by Teklemariam {\it et al.}~\cite{Teklemariam}.
They used a three-spin molecule as
a composite of a one-qubit system and a two-qubit environment.
Their initial state vector takes the form
$ | \chi_{\rm s} \ket \otimes 
  | \phi_{\rm e} \ket \otimes 
  | \psi_{\rm e} \ket $.
Their Hamiltonian in our notation is
\begin{eqnarray*}
	H 
	&=&
	J_{12} I_z \otimes I_z \otimes I
	+ J_{13} I_z \otimes I   \otimes I_z
	+ J_{23} I   \otimes I_z \otimes I_z,
\end{eqnarray*}
with which the time-evolution operator $ U(t) = e^{-iHt} $ is defined.
They introduced a kick operator
$$ 
	K ( \xi, \zeta ) 
	= I \otimes e^{i \xi \sigma_y} \otimes e^{i \zeta \sigma_y},
$$ 
which acts on the artificial environment.
Here $ \xi $ and $ \zeta $ are random variables,
which are interpreted as kick angles.
Now the time evolution operator of the whole system is 
\begin{eqnarray*}
	U_{\xi,\zeta}
	&=&
	K ( \xi_n, \zeta_n ) U(t)
	K ( \xi_{n-1}, \zeta_{n-1} ) U(t)
	\cdots
	\nonumber \\ &&
	\cdots
	K ( \xi_2, \zeta_2 ) U(t)
	K ( \xi_1, \zeta_1 ) U(t).
\end{eqnarray*}
This should be compared with our time-evolution operator (\ref{U_xi}).
Their strategy to manipulate the environment is different
from ours.
A channel associated with their model is defined
if $ U_{\xi,\zeta} $ is substituting into (\ref{channel}).
A two-qubit environment is required to
simulate an arbitrary one-qubit channel.
In contrast, we are interested only in the phase decoherence in this paper
and a one-qubit environment is sufficient for this purpose 
as was discussed above. It is possible, thanks to this simplification, to
carry out all the calculations analytically.

%=====================================================================
\section{Experiments}
%=====================================================================
\label{sec:experiment}

We demonstrate 
generation and suppression of phase decoherence experimentally
with an NMR quantum computer.
A 0.6 ml, 200 mM sample of ${}^{13}$C-labeled chloroform 
(Cambridge Isotope) in d-6 acetone is employed
as a two-qubit molecule.
The spin of carbon nucleus in chloroform is referred to as spin~1
(qubit~1), while
the spin of hydrogen nucleus is referred to as spin~2 (qubit~2).
Data is taken at room temperature 
with a JEOL ECA-500 NMR spectrometer~\cite{jeol}, whose 
hydrogen Larmor frequency is approximately 500~MHz. 
The measured spin-spin coupling constant is 
$ J / 2 \pi = 215.5$~Hz.
The transverse relaxation times in a natural environment are 
$ T_2 \sim 0.30 $~s for the carbon nucleus
and 
$ T_2 \sim 7.5 $~s for the hydrogen nucleus.
The longitudinal relaxation times are measured to be 
$ T_1 \sim 20 $~s for both nuclei. 
The duration of a $ \pi $-pulse for both nuclei is set to $ 50~\mu $s
throughout our experiments.  
Precision of pulse duration control is 100 ns.

\subsection{Hamiltonian in the rotating frame}
\label{rf_ham}
The approximate Hamiltonian 
of two spins in a heteronucleus molecule, 
such as $^{13}$C-labeled chloroform (two spins are $^{13}$C and H)
in a static magnetic field $B_0$ along the $z$-axis is 
\begin{equation}
	H =
	- \omega_{0,1} I_z \otimes I
	- \omega_{0,2} I   \otimes I_z
	+ J I_z \otimes I_z,
	\label{Hamiltonian}
\end{equation}
under the secular approximation \cite{spin_echo}.
Here $ \omega_{0,i} = \gamma_i B_0$,  
$ \gamma_i $ being the gyromagnetic ratio of the nucleus $ i $, and
$ J $ is a scalar coupling constant between the spins. 
The state of the whole system evolves following 
the Schr{\"o}dinger equation 
$ i \frac{d}{dt} | {\psi} (t) \ket = H | {\psi} (t) \ket $.
If we apply a time-dependent unitary transformation
\begin{equation}
	R = 
	e^{-i \omega_{0,1}  I_z t}
	\otimes 
	e^{-i \omega_{0,2} I_z t}
	\label{rotating frame}
\end{equation}
on the state of the two-qubit system as
$ | \tilde{\psi} (t) \ket = R | \psi (t) \ket $,
we obtain
\begin{eqnarray*}
	i \frac{d}{dt} | \tilde{\psi} (t) \ket 
	&=&
	 i R \frac{d}{dt} | {\psi} (t) \ket 
	+i  \frac{dR}{dt} | {\psi} (t) \ket 
\\
	&=&
	 R H | {\psi} (t) \ket 
	+i  \frac{dR}{dt} | {\psi} (t) \ket 
\\
	&=&
	 R H R^\dagger | \tilde{\psi} (t) \ket 
	+i  \frac{dR}{dt} R^\dagger | \tilde{\psi} (t) \ket.
\end{eqnarray*}
Therefore, the transformed state $ | \tilde{\psi} (t) \ket $
satisfies the Schr{\"o}dinger equation 
$ i \frac{d}{dt} | \tilde{\psi} (t) \ket = \tilde{H} | \tilde{\psi} (t) \ket $
with the transformed Hamiltonian
\begin{equation}
	\tilde{H}
	=
	R H R^\dagger +i \frac{dR}{dt} R^\dagger 
	=
	J I_z \otimes I_z ,
\end{equation}
which is nothing but the Hamiltonian (\ref{tilde H}). 
The time-dependent operator (\ref{rotating frame}) transforms
a coordinate system from the laboratory frame to a rotating frame.
We will use the rotating coordinate system defined with
Eq.~(\ref{rotating frame}) in the following
and the symbol $\tilde{}$ will be omitted henceforth to simplify the notations.

\subsection{Demonstration of the phase decoherence in a quantum transmission line}
\begin{figure}[t]
\begin{center}
\includegraphics[bb=0 0 370 400,width=7.5cm]{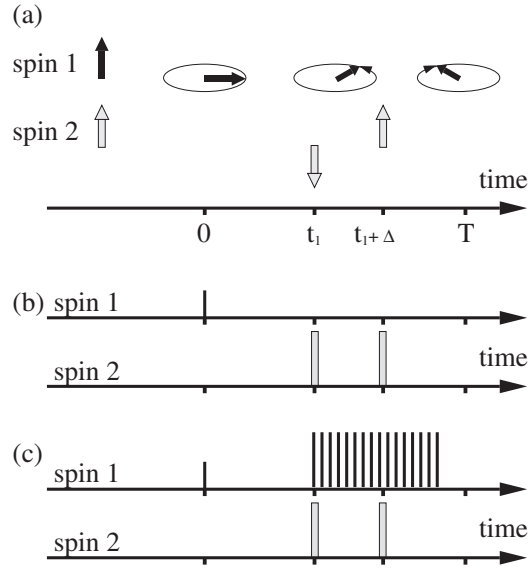}
\end{center}
\caption{
\label{fig2}
Experimental scheme to generate phase decoherence in 
a flying qubit in a transmission line. 
(a)~Motion of the spin Bloch vectors viewed in the rotating frame.
(b)~The pulse sequence to implement the phase flip channel.
A short bar denotes a $ \pi/2 $-pulse
while a long bar denotes a $ \pi $-pulse.
(c)~The pulse sequence for the bang-bang control.
A regular sequence of $ \pi $-pulses with a constant interval $ t_{\rm b} $
is applied on spin~1.
}
\end{figure}

The model of phase decoherence in a transmission line is implemented with NMR.
The scheme of the experiment is shown in Fig.~\ref{fig2}. 
Time evolution of spins is depicted 
in Fig.~\ref{fig2} (a) in terms of the
Bloch vectors viewed from the rotating frame defined with
the transformation~(\ref{rotating frame}).
Both spins are set initially
in the state $|0 \ket$.
This initial state is prepared 
as a so-called pseudopure state~\cite{pps}. 
Spin~1 is turned to the $x$-directon by a $ \pi/2 $-pulse along the $y$-axis
at $ t = 0 $.
Spin~2 is flipped 
by a $ \pi $-pulse at $ t = t_1 $
and then flipped back to $|0 \ket$ by another
$ \pi $-pulse at $ t = t_1 + \Delta $.
Spin~1 evolves
according to the Hamiltonians (\ref{H|0>}).
Spin~1 precesses with the angular velocity $ -J/2 \,(J/2)$ 
while spin~2 is in the state $|0 \rangle \, (|1\rangle)$.
The state of spin~1 at $ t = T $ is measured 
via a free induction decay (FID) signal.
Figure~\ref{fig2} (b) is a schematic picture 
of the pulse sequence to manipulate these spins.
A short bar corresponds to a $ \pi/2 $-pulse
while a long bar to a $ \pi $-pulse.

\begin{figure}[t]
\begin{center}
\includegraphics[bb=110 245 420 420,width=7.5cm]{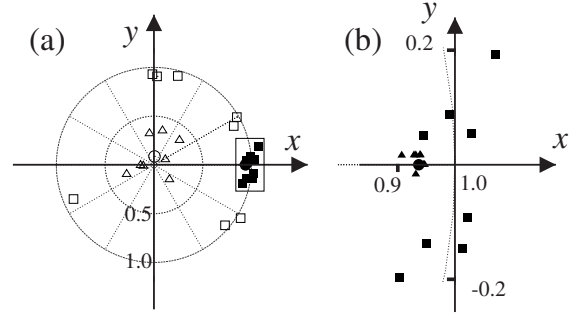}
\end{center}
\caption{
\label{fig3}
Amplitude of spin~1 measured 
in the presence of phase decoherence in a transmission line (open symbols)
and the same with the bang-bang control (filled symbols).
$ \square, \blacksquare $~: individual amplitude measured via FID signals. 
$ \triangle, \blacktriangle $~: amplitude averaged over 16 measurements.
$ \bigcirc, \bullet $~: amplitude averaged over 128 measurements 
($16 \times 8 =128$).
The rectangle area in~(a) is magnified in the panel~(b). 
}
\end{figure}

The measured quantities via FID signals are 
the components $ (a_x, a_y) $ of the Bloch vector of spin~1,
which correspond to the off-diagonal elements
of the density matrix (\ref{parameters of rho}).
The measured complex amplitudes 
$ \{ a_x + i a_y = e^{i \phi} \sqrt{ a_x^2 + a_y^2 } \} $
are plotted in Fig.~\ref{fig3}.
The measurements are repeated with
variety of spin~2 inversion time $\Delta$ uniformly distributed
in the range $ [0, 2\pi/J] =[0, 1/215.5]~$s.
The total number of repetition is 128 in our experiment.

The open squares~$ \square $ in Fig.~\ref{fig3}
show the measured amplitudes, where only 8 out of 128 amplitudes are shown.
Their absolute values are close to unity while
their phases are distributed in the range $[0, 2\pi]$
due to the variation of spin~2 inversion time $\Delta$. 
The contribution of the trivial phase shift $ S(-{JT}/{2}) $ 
in Eq.~(\ref{eq:ps_T}) has been removed when plotting the data.

To construct a mixing process we take average of measured amplitudes.
Each open triangle~$ \triangle $ in Fig.~\ref{fig3}
is an average of 16 amplitudes.
There are $8 =128/16$ open triangles in total.
It is found that the absolute values 
of the averaged amplitudes~$ \triangle $ become 
considerably smaller than unity.

The open circle~$ \bigcirc $ in Fig.~\ref{fig3} 
shows the average of all 128 measured amplitudes.
The averaged amplitude is close to the origin, which
implies vanishing off-diagonal elements of the density matrix
of spin~1
and therefore is a clear indication of phase decoherence.
Thus we see that 
this system works as an artificial phase flip channel for spin~1.

Next, we apply the bang-bang control to the system qubit.
Figure~\ref{fig2} (c) shows the pulse sequence to implement
the phase flip channel and the bang-bang control.
We start applying a regular series of short $ \pi $-pulses on spin~1
at $t=t_1$, whose pulse interval is $ t_{\rm b} = 0.3$~ms.
The number of $ \pi $-pulses is 16 in each run.
The duration of the $ \pi $-pulses is a sum of 
the pulse intervals (0.3~ms) $\times 15$ 
and the pulse durations (50~$\mu$s) $\times 16$
and thus is 5.3~ms, which covers
the period when spin~2 is in the state $|1\ket$.
The rotation axes of the $ \pi $-pulses for spin~1 
are cyclically permutated as 
\begin{equation}
(x, -x, y, -y, -x, x, -y, y)
\label{eq:cycle}
\end{equation} 
in order to reduce influence of possible pulse imperfections. 
We repeat this procedure 128 times
with randomly varied $ \Delta $'s.

The FID amplitudes in the presence of the bang-bang pulses
are indicated by filled squares~$ \blacksquare $ in Fig.~\ref{fig3}.
The panel~(b) in Fig.~\ref{fig3} 
is a magnification of a part of the panel~(a).
The absolute values of these amplitudes are almost one.
Moreover, their phases are concentrated in a narrow range,
$ | \phi | \le 0.25$~rad. This is comparable with the theoretical
estimate 
$ J t_{\rm b} = 2 \pi \times 215.5 \times 0.3\times 10^{-3} \simeq 0.4 $~rad.
The filled triangle~$ \blacktriangle $ in Fig.~\ref{fig3}
shows the average of 16 amplitudes in the presence of the bang-bang control.
The filled circle~$\bullet $ in Fig.~\ref{fig3}
shows the average of whole 128 amplitudes.
The average of the whole data is $ 0.93+0.0\, i \pm 0.1 \, i $.
The error in the $ y $-component 
originates form the FID phase determination error. 

This magnitude 0.93 may be compared 
with the theoretical prediction of 0.99. 
Note that we should slightly modify $\kappa $ from 
Eq.~(\ref{kappa}) to 
$\displaystyle \kappa = \frac{ \sin (J t_{\rm b}/2)}{ J t_{\rm b}/2 } , $
since the starting times of the bang-bang
pulses are fixed at $t = t_1$ in experiments. 
This discrepancy between our theory 
and experiments should not be taken seriously
since the pulse durations (50~$\mu$s) for qubit~1 and ~2 
are finite and are 1/6 of the pulse interval (0.3~ms) of the 
bang-bang pulses in real experiments while they are 
assumed infinitely short in theory.  

We conclude this section by stating that the bang-bang pulses 
really suppress decoherence generated artificially.

\subsection{Demonstration of the phase decoherence in a quantum memory}

The model of the phase decoherence in a quantum memory is also implemented 
with NMR.
We use the same two-spin molecule as that employed in the previous experiment.
Both spins are initially set to the up-state $|0 \ket$.
Spin~1 is turned to the $x$-axis by a $ \pi/2 $-pulse at $ t = t_0 = 0 $ 
while
spin~2 is flipped from $|0 \ket$ to $|1 \ket$
by a $ \pi $-pulse at $ t = t_{2k-1} $
and is flipped back from $|1 \ket$ to $|0 \ket$ by a 
subsequent $ \pi $-pulse at $ t = t_{2k} $ $ (k=1,2,3, \cdots) $.
The rotation axes of these $ \pi $-pulses
are cyclically permutated as in Eq.~(\ref{eq:cycle})
to reduce undesired influence of imperfections in the $ \pi $-pulses.
Thus the number of $\pi$-pulses are set to a multiple of 8.
The time intervals $ \{ \Delta_j = t_{j+1} - t_{j} \} $ between
adjacent $\pi$-pulses distribute
according to the Gaussian distribution
(\ref{random interval}) and (\ref{Gaussian}).
The spin~1 evolves with the Hamiltonians (\ref{H|0>}).
The $x$- and $y$-components of the Bloch vector of spin~1 at $ t = T $ is 
measured via a FID signal.
%The magnitude $ \sqrt{a_x^2 + a_y^2} $ of the measured amplitude 
%is plotted as a function of $ T $ in Fig.~\ref{fig5}.

\begin{figure}[b]
\begin{center}
\includegraphics[bb=90 140 480 450,width=7.5cm]{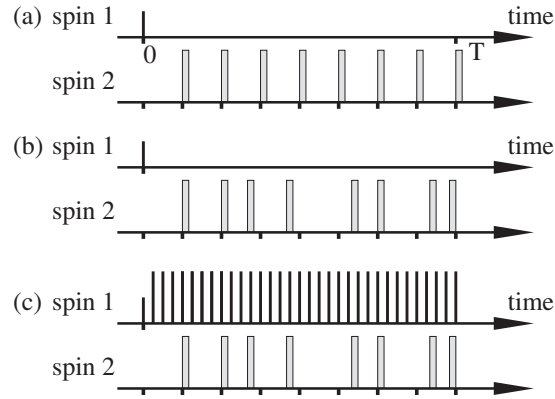}
\end{center}
\caption{
\label{fig4}
Pulse sequences to implement
artificial phase decoherence in a quantum memory.
Both spins are initially in the up-states.
A short bar is a $ \pi/2 $-pulse
while a long bar is a $ \pi $-pulse.
(a)~A regular series of pulses with a fixed interval $ \bar{\Delta} = 2.0 $~ms
is applied on spin~2 as a reference to other experiments.
(b)~A series of pulses with random intervals is applied on spin~2.
The variance $\alpha$ of pulse intervals is adjustable.
(c)~The bang-bang control pulses are applied on spin~1 
while random pulses are applied on spin~2.
}
\end{figure}

\begin{figure}[b]
\begin{center}
\includegraphics[bb=120 50 500 360,width=8cm]{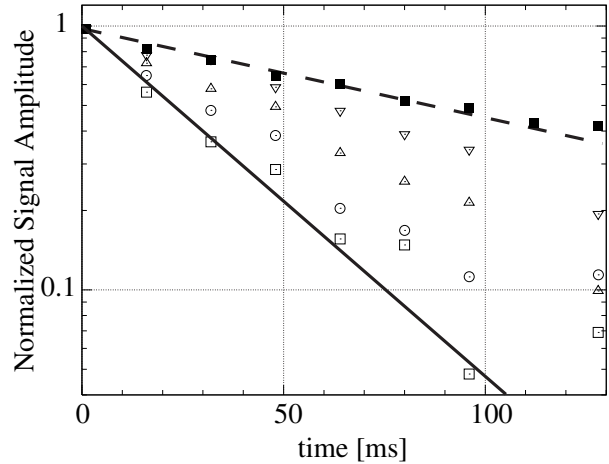}
\end{center}
\caption{
\label{fig5}
Time evolution of the magnitude $ |a_x + i a_y| $ of spin~1.
Decrease in these magnitude indicates phase decoherence.
The correspondence between the symbol and the variance $\alpha$ of the 
pulse interval distribution is
$ ( \blacksquare : \, \alpha= 0 ) $,
$ ( \bigtriangledown : \, \alpha= 0.10 ) $,
$ ( \triangle : \, \alpha= 0.15 ) $,
$ ( \bigcirc : \, \alpha= 0.20 ) $ and
$ ( \square : \, \alpha= 0.25 ) $.
The broken line is the least square fit for $ \{ \blacksquare \} $ while
the solid line is % the least square fit 
that for $ \{ \square \} $.
}
\end{figure}

In the first run, depicted in Fig.~\ref{fig4}~(a),
we put $ \alpha = 0 $ and hence
the time interval between $\pi$-pulses is a constant,
$ \Delta_j = \bar{\Delta} = $ 2.0~ms. 
In this case 
a regular alternating field acts on spin~1.
If the $\pi$-pulses and the spin dynamics were perfect, 
there would be no decoherence.
However, in reality, it is impossible to avoid pulse imperfection,
measurement errors and intrinsic decoherence.
Figure~\ref{fig5} shows that
decoherence takes place even in the system under regular pulses.
The measurement with $ \alpha = 0 $ is necessary as a reference
to the other measurements with $ \alpha \ne 0 $. 
We may claim that decoherence is enhanced by the random fluctuating field
if we observe faster decoherence in the measurement with $ \alpha \ne 0 $
than that with $ \alpha = 0 $.
The magnitudes of measured amplitudes under the pulses with $ \alpha = 0 $
are plotted as filled squares~$ \blacksquare $ in Fig.~\ref{fig5}.
The measured decoherence factor for $\alpha=0$ is 
\begin{equation}
	e^{- T / T_2^*} = 0.45
	\qquad \mbox{at $T=100$~ms.}
	\label{reference data}
\end{equation}

Next, we modulate the time interval of spin~2 flippings randomly.
The corresponding pulse sequence is shown in Fig.~\ref{fig4}~(b).
The variance of the intervals in Eq.~(\ref{random interval}) 
is chosen to be
$ \alpha = $ 0.10, 0.15, 0.20 and 0.25.
The average of the intervals is $ \bar{\Delta} = 2.0 $~ms.
A series of random variables $ \Xi =( \xi_0, \xi_1, \cdots, \xi_r ) $
is prepared according to the Gaussian distribution (\ref{Gaussian}).
Then a series of intervals 
$ (\Delta_0, \Delta_1, \cdots, \Delta_r ) $
is defined via (\ref{random interval}) and then
the amplitude $ a_x + i a_y $ of spin~1 is measured at time $ t = T $.
We prepare 128 series $ \{ \Xi_1, \Xi_2, \cdots, \Xi_{128} \} $,
repeat measurements 
128 times and take an average of 128 measured amplitudes
for each values of $ \alpha $ and $ T $.
The magnitude of the averaged amplitude is plotted as
a function of $ T $ in Fig.~\ref{fig5}.
The correspondence between the symbol and the 
variance parameter $\alpha$ is
$ ( \bigtriangledown : \, \alpha= 0.10 ) $,
$ ( \triangle : \, \alpha= 0.15 ) $,
$ ( \bigcirc : \, \alpha= 0.20 ) $,
$ ( \square : \, \alpha= 0.25 ) $.
The plotted data show
exponential decrease in the magnitude of the amplitude
$a_x + i a_y$.
It is clearly seen that a larger variance $\alpha$ 
introduces faster decoherence in spin~1.
The decoherence factors are read from 
the slopes of the graphs in Fig.~\ref{fig5} as
\begin{eqnarray}
	e^{- T / T_2^*} 
	&=&
	0.31, \;
	0.20, \;
	0.10, \;
	0.05
	\quad
	{\rm at} \; T = 100 \, \rm{ms} \;\; 
\nonumber \\ 
	{\rm for} \; 
	\alpha &=& 
	0.10, \;
	0.15, \;
	0.20, \;
	0.25.
	\label{data}
\end{eqnarray}

Let us introduce a dimensionless quantity
\begin{equation}
	R (\alpha) =
	- \frac{1}{\alpha^2}
	\ln \left\{
		\frac{ e^{- T / T_2^*} |_{ \alpha \ne 0} }
		{ e^{- T / T_2^*} |_{ \alpha = 0} }
	\right\}
	\label{Q}
\end{equation}
to compare the measured data with the theoretical estimation.
The theoretical prediction (\ref{prediction of random fluc}) yields
\begin{equation}
	R_{\rm theory} 
	= \frac{1}{8} J^2 \bar{\Delta} \, T
	= 46
	\label{theoretical estimation}
\end{equation}
for $ J = 2 \pi \times 215.5 $~Hz, 
$ \bar{\Delta} = 2 $~ms,
$ T = 100 $~ms, independently of 
% The theory also predicts that $ Q(\alpha) $ is independent from 
$ \alpha $.
On the other hand, by substituting
the measured values 
(\ref{reference data}) and (\ref{data}) into (\ref{Q}), we obtain
\begin{eqnarray}
	\label{measured}
\begin{array}{cccccc}
	R (\alpha)          &=  &37,  &36,   &38,   &35
\\
	{\rm for} \; \alpha &= &0.10, &0.15, &0.20, &0.25.
\end{array}
\nonumber
\end{eqnarray}
We observe that
the value $ R (\alpha) $ is almost independent of $ \alpha $,
implying that the decoherence rate $ (T_2^*)^{-1} $
is proportional to $ \alpha^2 $ as predicted 
in~(\ref{prediction of T2}).

We conducted a numerical simulation and found
that calibration error and spatial inhomogeneity of 
rf pulse fields may lead to apparently longer $T_2^*$
than that with perfect $\pi $-pulses. Therefore, 
we attribute the small quantitative discrepancy 
between the theory and the experiments to rf pulse imperfections.   
It is observed in Fig.~\ref{fig5}
that the data points deviate from the straight line
in particular for the case with 
$ \alpha = 0.25 $ with $ T \ge 80 $~ms.
Our numerical simulation also shows that
fluctuation in averaged amplitude is large in the
region where the averaged amplitude is small
This fluctuation originates from smallness of the statistical ensemble, 
whose size is 128 in our experiment.

\begin{figure}[b]
\begin{center}
\includegraphics[bb=90 20 480 340, width=8cm]{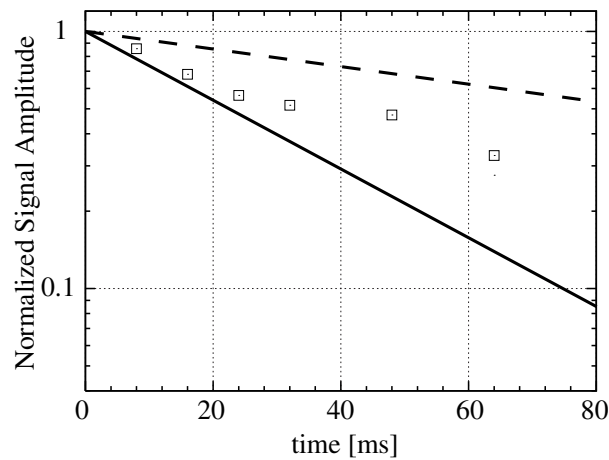}
\end{center}
\caption{
\label{fig6} 
Magnitudes $|a_x+i a_y|$ of spin~1 with the bang-bang control 
($t_{\rm b} = 0.5$~ms) applied are shown with the symbol $( \square)$.
The broken line and the solid line are the same
as in Fig.~\ref{fig5}. Series of $ \pi $-pulses 
with variance $ \alpha = 0.25 $ is applied to spin~2.
}
\end{figure}

Next, we apply the bang-bang control to spin~1.
The pulse sequence to incorporate the bang-bang control
is shown in Fig.~\ref{fig4}~(c). 
A regular sequence of $ \pi $-pulses with interval 
$ t_{\rm b} = 0.5$~ms is applied on spin~1.
The rotation axes of these $ \pi $-pulses
are cyclically permutated as given in Eq.~(\ref{eq:cycle}).
During this run,
a sequence of $ \pi $-pulses whose interval fluctuates
with variance $ \alpha = 0.25 $
is also applied on spin~2.
We finally measure the amplitude of spin~1 at $ t = T $.
We repeat the measurement by preparing 128 series of variables
$ \{ \Xi_1, \Xi_2, \cdots, \Xi_{128} \} $ as input parameters.
The magnitude of averaged amplitude is plotted 
as a function of $ T $ with the symbol 
$ ( \square ) $ in Fig.~\ref{fig6}. 
The broken line in Fig.~\ref{fig6} is the reference data
with regular pulses on spin~2 and no pulses on spin~1 (Fig.~\ref{fig4}~(a)).
The solid line in Fig.~\ref{fig6} is 
the experimental result with random pulses on spin~2
and without the bang-bang control to spin~1 (Fig.~\ref{fig4}~(b)).
Comparing the data points~$ \{ \square \} $ with the solid line,
we observe that decoherence is suppressed by the bang-bang control.
The decoherence factor is read 
from $ \{ \square \} $ in Fig.~\ref{fig6} as
\begin{equation}
	e^{ - T/T_2^* } = 0.35
	\;\;
	{\rm at} \; T = 60 \, {\rm ms} \;\; 
	{\rm for} \; t_{\rm b} = 0.50 \, {\rm ms}.
\end{equation}
The decoherence factor of the reference broken line in Fig.~\ref{fig6} is
\begin{equation}
	e^{ - T/T_2^* } = 0.63
	\;\;
	{\rm at} \; T = 60 \, {\rm ms} \;\; 
	{\rm for} \; \alpha = 0.
\end{equation}
Their ratio is
\begin{equation}
	\frac{ \; e^{-T/T_2^*} |_{\rm bang} \; }{ e^{-T/T_2^*} |_{\rm ref}}
	= \frac{0.35}{0.63}
	= 0.56.
	\label{observed ratio}
\end{equation}
%On the other hand, 
Theoretical prediction (\ref{bang-bang prediction}) leads
\begin{equation}
	e^{ - T/T_{2 {\rm b}}^* } |_{\rm theory} = 0.56,
\end{equation}
which is in good agreement with experiments,
although there are various factors not taken into account in our theory. 
For example, 
finite pulse durations, pulse calibration errors, and 
adequacy of the assumption of $J t_{\rm b} \ll 2\pi$ are ignored in our
theoretical analysis. 

From these results we conclude that
the bang-bang pulses have suppressed decoherence caused 
by the interaction with the dynamical environment. %, see $\square$'s. 

%=====================================================================
\section{Conclusions}
%=====================================================================
\label{sec:conclusion}

We have shown in this paper
that a two-qubit system can simulate a composite of 
a system (qubit~1) and its environment (qubit~2) so that qubit~1
exhibits phase decoherence, provided that
the state of qubit~2 is randomized by external manipulation. 
We have evaluated decoherence rates of qubit~1 and also shown 
that decoherence is suppressed by applying the bang-bang control to 
qubit~1.

Performing NMR experiments with a two-spin molecule
we measured decoherence in a clear manner.
In the simulation of phase decoherence in a qubit flying through 
a quantum transmission line, our theoretical calculations
were consistent with the measured amplitudes.
Our theoretical calculations
qualitatively explained the measured decoherence rates
in the simulation of phase decoherence in a quantum memory.
It was confirmed that
the decoherence rate $ (T_2^*)^{-1} $ is proportional to the squared variance
$ \alpha^2 $ of the interval distribution of the pulses applied
to the environment (qubit~2).
In both cases we demonstrated that the bang-bang control successfully 
suppressed decoherence when the interval $t_{\rm b}$
of successive time reversal 
operations is much shorter than the correlation time of 
the artificial environment.

Study of a qubit system as a composite of a system and 
its environment will help our understanding of the mechanism 
of decoherence and 
will help further development of techniques
to suppress decoherence.

\section*{Acknowledgments}

We would like to thank Paolo Zanardi for giving us a series of lectures on 
quantum theory of open systems.
MN would like to thank Mikko Paalanen and Jukka Pekola for warm
hospitality extended to him during his stay at Low Temperature Laboratory,
Helsinki University of Technology, Finland, where a part of this work was 
done. MN is partially supported by MEXT, Grant No.~13135215.
ST is partially supported by JSPS, Grant Nos.~15540277 and 17540372.


\begin{thebibliography}{99} 
\bibitem {ref:1} 
         M.\ A.\ Nielsen and I.\ L.\ Chuang,
         {\it Quantum Computation and Quantum Information} 
         (Cambridge University Press, Cambridge, 2000).

\bibitem {VSB01} 
         L.\ M.\ K.\ Vandersypen, 
         M.\ Steffen, G.\ Breyta,
         C.\ S.\ Yannoni, M.\ H.\ Sherwood, and I.\ L.\ Chuang, 
         Nature {\bf 414}, 883 (2001).
% Experimental realization of Shor's quantum factoring algorithm 
% using nuclear magnetic resonance 

\bibitem {pt}
         R.\ Fitzgerald, Physics Today, {\bf 53}, No.\ 1, 20 (2000).

\bibitem {qaa} 
         M.\ Nakahara, Y.\ Kondo, K.\ Hata, and S.\ Tanimura, 
         Phys.\ Rev.\ A {\bf 70}, 052319 (2004).
\bibitem {warp}
         M.\ Nakahara, J.\ J.\ Vartiainen, Y.\ Kondo,
         S.\ Tanimura, and K.\ Hata, 
         Phys.\ Lett.\ A {\bf 350}, 27 (2006).
%         to be published. See also arXiv: quant-ph/0411153. 

\bibitem{Zanardi}
	P. Zanardi,
	Phys. Rev. A {\bf 56}, 4445 (1997);
%	Dissipative dynamics in a quantum register
	Phys. Rev. A {\bf 57}, 3276 (1998).
%	Dissipation and Decoherence in a Quantum Register

\bibitem {bb}
         M.\ Ban, J.\ Mod.\ Opt.\ {\bf 45}, 2315 (1998);
         L.\ Viola and S.\ Lloyd, Phys.\ Rev.\ A {\bf 58}, 2733 (1998);
%        % Lu-Ming Duan and Guang-Can Guo,
         L.\ Duan and G.\ Guo, Phys.\ Lett.\ A {\bf 261}, 139 (1999);
         L.\ Viola, E.\ Knill, and S.\ Lloyd, Phys.\ Rev.\ Lett.\ {\bf 82}, 
         2417 (1999);
         L.\ Viola, S.\ Lloyd, and E.\ Knill, Phys.\ Rev.\ Lett.\ {\bf 83}, 
         4888 (1999);
         L.\ Tian and S.\ Lloyd, Phys.\ Rev.\ A {\bf 62}, 050301(R) (2000).
         See, also, H.\ Gutmann, F.\ K.\ Wilhelm, 
         W.\ M.\ Kaminsky, and S.\ Lloyd, 
         {\it Bang-Bang Refocusing of a Qubit Exposed to Telegraph Noise} 
         in 
         {\it Experimental Aspects of Quantum Computing}, 
         edited by H.\ O.\ Everitt 
         (Springer, New York, 2005).

\bibitem {uchiyama}
         C.\ Uchiyama and M.\ Aihara, 
         Phys.\ Rev.\ A {\bf 66}, 032313 (2002); 
%         C.\ Uchiyama and M.\ Aihara, 
         Phys.\ Rev.\ A {\bf 68}, 052302 (2003).

\bibitem{Hellwig}
	K.-E. Hellwig and K. Kraus, 
	Commun. Math. Phys. {\bf 11}, 214 (1969).
%	``Pure operations and measurements''

\bibitem{Kraus}
	K. Kraus, 
	Ann. Phys. {\bf 64}, 311 (1971).
%	``General state changes in quantum theory''

\bibitem{ECGS61}
        E.\ C.\ G.\ Sudarshan, P.\ M.\ Mathews, and J.\ Rau, 
        Phys.\ Rev.\ {\bf 121}, 920 (1961).

\bibitem {Leung99}
         D.\ Leung, L.\ Vandersypen, X.\ Zhou, M.\ Sherwood, C.\ Yannoni,
         M.\ Kubinec, and I.\ Chuang, Phys.\ Rev.\ A {\bf 60}, 1924 (1999).
%        Experimental realization of a two-bit phase damping quantum code

\bibitem{ref:rff}
	P.\ W.\ Anderson, J.\ Phys.\ Soc.\ Jpn.\ {\bf 9}, 316 (1954);
	R.\ Kubo, J.\ Phys.\ Soc.\ Jpn.\ {\bf 9}, 935 (1954);
	R.\ Kubo, 
%	A Stochastic Theory of Line-Shape and Relaxation, 
	in {\it Fluctuation, Relaxation and Resonance in Magnetic Systems}, 
	ed. by ter Haar (Oliver and Boyd, Edinburgh, 1962);
	R.\ Kubo and N.\ Hashitsume, 
	{\it Statistical Physics II, Nonequilibrium Statistical Mechanics} 
	(Springer-Verlag, Berlin, Heidelberg, New York, Tokyo, 1985).

\bibitem {zurek}
         W.\ H.\ Zurek, Phys.\ Rev.\ D {\bf 26}, 1862 (1982).

\bibitem {zhang}
         J.\ Zhang, 
         Z.\ Lu, L.\ Shan, and Z.\ Deng,
         arXiv: quant-ph/0202146; 
%	Experimental study of quantum decoherence using nuclear magnetic resonance
%       J.\ Zhang, %{\it et al},
%       Z.\ Lu, L.\ Shan, and Z.\ Deng, arXiv: 
        quant-ph/0204113. 
%	Simulating decoherence behavior of a system in entangled state 
%	using nuclear magnetic resonance

\bibitem {ryan}
        C.\ A.\ Ryan, J.\ Emerson, D.\ Poulin, C.\ Negrevergne, 
        and R. Laflamme, 
        Phys.\ Rev.\ Lett.\ {\bf 95}, 250502 (2005). 
%       Characterization of Complex Quantum Dynamics 
%       with a Scalable NMR Information Processor

\bibitem {Teklemariam}
         G.\ Teklemariam, 
         E.\ M.\ Fortunato, C.\ C.\ L\'{o}pez, J.\ Emerson, 
         J.\ P.\ Paz, T.\ F.\ Havel, and D.\ G.\ Cory,
         Phys.\ Rev.\ A {\bf 67}, 062316 (2003).
%	Method for modeling decoherence on a quantum-information processor

\bibitem {ernst} 
         R.\ R.\ Ernst, J.\ Chem.\ Phys.\ {\bf 45}, 3845 (1966). 

\bibitem{Morton}
         J.\ J.\ L.\ Morton, A.\ M.\ Tyryshkin, A.\ Ardavan, 
         S.\ C.\ Benjamin, K.\ Porfyrakis, S.\ A.\ Lyon, 
         and G.\ A.\ D.\ Briggs, Nature Physics {\bf 2}, 40 (2006). 

\bibitem {Itano} 
         W.\ M.\ Itano, D.\ J.\ Heinzen, J.\ J.\ Bollinger, 
         and D.\ J.\ Wineland, Phys.\ Rev.\ A {\bf 41}, 2295 (1990). 

\bibitem{Cook}
        R.\ J.\ Cook, Phys.\ Scr.\ T {\bf 21}, 49 (1988).
         
\bibitem {Ref-FK}
         T.\ Kohmoto, Y.\ Fukuda, M.\ Kunitomo, 
         K.\ Ishikawa, Y.\ Takahashi, K.\ Ebina, and M.\ Kaburagi,
         Phys. Rev. B {\bf 52}, 13475 (1995). % 13475-13479
%	Hole burning in a well-characterized noise field: 
%       Nonadherence to the Bloch equations

\bibitem {NS}
         L.\ Viola, E.\ M.\ Fortunato, M.\ A.\ Pravia, E.\ Knill, 
         R.\ Laflamme, and D.\ G.\ Cory, Science {\bf 293}, 2059 (2001).    

\bibitem {DFS}
         P.\ G.\ Kwiat, A.\ J.\ Berglund, J.\ B.\ Altepeter, 
         and A.\ G.\ White, Science {\bf 290}, 498 (2000). 

\bibitem {ion}
         D.\ Kielpinski, V.\ Meyer, M.\ A.\ Rowe, C.\ A.\ Sackett, 
         W.\ M.\ Itano, C.\ Monroe, and D.\ J.\ Wineland, 
         Science {\bf 291}, 1013 (2001). 
   
\bibitem {kitajima}
         S.\ Kitajima, M.\ Ban, and F.\ Shibata, No.\ 25aYG12 
         in the 60th annual meeting of the Physical Society of Japan (2005).  

\bibitem {random_walk}
         D.\ Pines and P.\ Slichter, Phys.\ Rev.\ {\bf 100}, 1014 (1955).

\bibitem {jeol}
         %http://www.jeol.co.jp/, 
         http://www.jeol.com/. 

\bibitem {spin_echo}
         M.\ H.\ Levitt, {\it Spin Dynamics} (John Wiley and Sons, New
         York, 2001). 

\bibitem {pps} 
         U.\ Sakaguchi, H.\ Ozawa, and T.\ Fukumi, 
         Phys.\ Rev.\ A {\bf 61}, 042313 (2000).
%        Method for effective pure states with any number of spins

\end{thebibliography}
\end{document}